# Electrical Route to Realising Intensity Simulation of Heavy Rain Events in Tropics

Dipjyoti Mudiar[1,2], Anupam Hazra[1], S. D. Pawar[1], Rama Krishna Karumuri[1], Mahen Konwar[1], Subrata Mukherjee[1], M. K. Srivastava[2] and B. N. Goswami[3]

[1]Indian Institute of Tropical Meteorology, Pune, India 411008

[2]Banaras Hindu University, Varanasi, India 221005

[3]Cotton University, Guwahati, India 781001

**Abstract:**

In the backdrop of a revolution in weather prediction by Numerical Weather Prediction (NWP) models, quantitative prediction of intensity of heavy rainfall events and associated disasters has remained a challenge. Encouraged by compelling evidence of electrical influences on cloud/rain microphysical processes, here we propose a hypothesis that modification of raindrop size distribution (RDSD) towards larger drop sizes through enhanced collision-coalescence facilitated by cloud electric fields could be one of the factors responsible for intensity errors in weather/climate models. The robustness of the hypothesis is confirmed through a series of simulations of strongly electrified (SE) rain events and weakly electrified (WE) events with a convection-permitting weather prediction model incorporating the electrically modified RDSD parameters in the model physics. Our results indicate a possible roadmap for improving hazard prediction associated with extreme rainfall events in weather prediction models and climatological dry bias of precipitation simulation in many climate models.

1. **Introduction**

The revolution in weather forecasting (*Boer et al.*, 2014) has led to significant improvement of simulation of precipitation in synoptic and mesoscales by Numerical Weather Prediction (NWP) models. However, the quantitative precipitation forecast (QPF) on a smaller scale, required for hydrological forecasts remains a challenge even in the latest high resolution operational models (*Shrestha et al.*, 2013; *Wang et al.*, 2016; *Shahrban et al.*, 2016) with unacceptably large mean absolute error, MAE (*Giinaros et al.*, 2015). The problem of errors in the QPF appears to be related to (a) displacement of the simulated centre of the mesoscale system compared to observed, (b) simulation of the phase of the diurnal cycle of precipitation by models a few hours before observed over land (*Dirmeyer et al.*, 2012) and (c) underestimation of heavy precipitation by almost all climate models even up to resolution of 12 km (*Kendon et al.*, 2012). As the same factors are also responsible for prediction errors of thunderstorms and extreme rainfall events, it is critical to improve them in models for skilful predictions of hazards associated with increasing frequency of extreme rainfall events (*Goswami et al.*, 2006). While there is a need for improving all three aspects of precipitation simulation in a model, in this study we focus only on the 'intensity' simulation of a convection-permitting NWP model. A simple increase in resolution of a model, however, is not helpful as has been found that it has little impact on the skill of prediction (*Shrestha et al.*, 2013) or produces too intense extreme events (*Kendon et al.*, 2012). It is recognized that high 'resolution' in a climate model is a necessary but not sufficient condition for simulating the variance of high-frequency fluctuations (*Goswami and Goswami*, 2016). It is also known that an adequate 'cloud microphysics' parameterization is essential for simulation of the organization of mesoscale systems and equatorial waves (*Hazra et al.*, 2017, 2019). However, numerical simulation of electrical forces within clouds associated with extreme rainfall events are only beginning to be addressed in NWP models (*Dafies et al.*, 2018). Here, we test a hypothesis that a large part of underestimation of the 'intensity' may be related to modification of the raindrop size distribution (RDSD) by electric fields in the clouds and test the veracity of the hypothesis through simulations of rainfall in several 'strongly' electrified cases and 'weakly' electrified cases in a convection-permitting NWP model.

A substantial fraction of tropical precipitation (57-60%) originates from thunderstorms and electrified shower cloud, the cloud with stronger in-cloud electrical environment but not substantial enough to produce lightning associated with Mesoscale Convective Systems (MCS), both of which exhibit strong electrical environment (*MacGorman et al*., 2008; *Liu et al*., 2010). Two dynamically and micro-physically distinct cloud regimes observed to contribute to the rainfall in the tropical atmosphere, i.e., convective and stratiform (*Houze*, 1997). In convective regime, where air updraft is stronger, the precipitation particles grow by accretion of cloud liquid water, a process known as *coalescence* in the warm phase of cloud and *riming* in the mixed phase (*Houghton*, 1968). In the stratiform regime, where there is not much liquid water present, the precipitation particles primarily grow by vapour diffusion and aggregation above the freezing layer of cloud (*Houze,* 1997). Apart from the prevailing dynamics and microphysics, the precipitation formation processes are known to influenced by the ambient aerosol size distribution as well (*Khain et al*. 1999; *Rosenfeld*, 2000; *Rosenfeld et al*. 2002; *Rosenfeld and Woodley*, 2003), although the relationship between them is observed to be non-linear. The scientific speculation regarding the electrical influence on the cloud microphysical processes is long-standing, dated back to the time of *Lord Rayleigh*, (1879). The lightning-producing clouds exhibit stronger in-cloud electrical environment with vertical electric field reaching values up to 400 kVm$^{-1}$ (*Winn et al.,* 1974) with charge densities which could go up to $10^9$ elementary charges (*Christian et al*.,1980; *Bateman et al*.,1999). Several laboratory, observational and numerical modeling studies provide compelling evidence suggesting strong electrical influences on cloud/rain microphysical processes inside strongly electrified cloud (*Schlamp et al*.,1976, 1979; *Khain et al*. 2004; *Bhalwankar and Kamra*. 2007; *Hortal et al*. 2012; *Harrison et al*.,2015). Numerical calculation of collision efficiency between two charged cloud droplets in an external electric field (vertically upward and downward) by *Schlamp et al.* (1976, 1979) and *Khain et al*. (2004) reported a significant effect of an external electric field and electric charges residing on the interacting drops on the collision efficiency of the drops. A few laboratory studies also revealed that presence of a vertical electric field can broaden the rain RDSD and hence enhance the growth rate of raindrops (*Bhalwankar and Kamra*, 2007). Laboratory investigation of *Ochs and Czys* (1987) reported that permanent coalescence results for all impact angles upon collision of two drops if their relative charge exceeds $2\times10^{-12}$ C irrespective of the polarity of the charges they carry. Our recent work of simultaneous field observations of RDSD and electrification of clouds at the High Attitude Cloud Physics Laboratory (HACPL), India (*Mudiar et al*. 2018) strongly supports some of the earlier laboratory and modeling studies. The observed similarity in the growth rate of raindrops in the warm phase of stratiform (*Mudiar et al.*, 2018) and convective strongly electrified cloud (*Mattos et al.*, 2016) indicates significant influence of electric force on coalescence growth of raindrops which evidently differentiate the evolutionary track of raindrops in strongly electrified cloud from weakly electrified ones.

Our hypothesis proposed above has emerged from these compelling and consistent evidence of the electrical influences on the cloud microphysical processes indicating the urgent need to include the electrical effects in the rain formation process in NWP models. Intrigued by this possibility, here we have attempted to test an NWP model's fidelity in simulating 8 rain events associated with a stronger in-cloud electrical environment and 5 rain events with weaker electric environment using the same model setup. The simulated precipitation fields are compared with the available observed data for validation. Attempts have also been made to bring in the electrical influences to the model physics schemes through modification of model RDSD parameters.

1. Data & Methodology

All the simulations pertaining to the current study were performed using Advanced Weather Research and Forecasting (WRF-ARW) model version 3.5.1 developed by the National Centre for Atmospheric Research (NCAR). The WRF is fully compressible, non-hydrostatic, terrain-following 3D mesoscale model. The simulations are performed considering four nested domain d01, d02, d03, d04 with a horizontal grid spacing of 27km, 9km, 3km &1km respectively. Figure 1a shows the geographical coverage of the model domain along with the topographical map (Figure 1b) of the innermost domain.

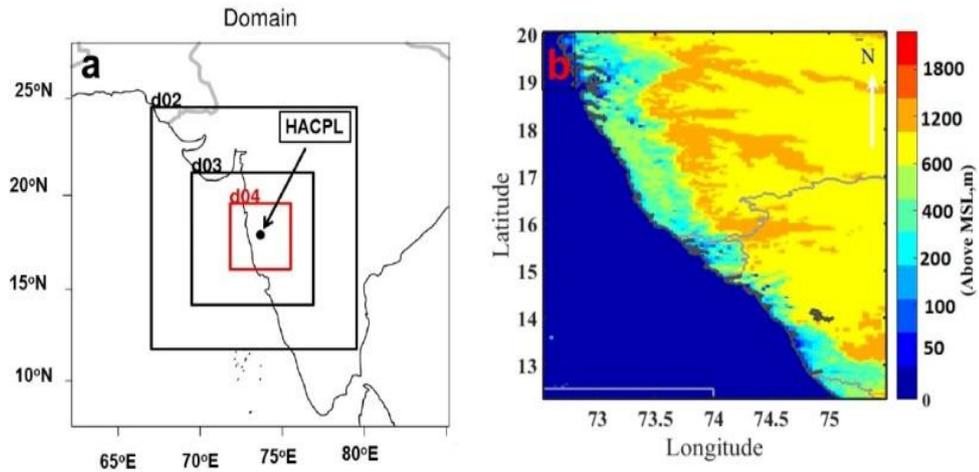

**Figure 1:** (a) Nested model domain, (b) topographical map encompassing domain d04.

The innermost domain d04 is centre at the HACPL, Mahabaleshwar, (India; 17.92 N, 73.66 E). The initial and boundary conditions are provided from 6 hourly National Centre for Environmental Prediction (NCEP) Final operational global analysis data with $1°\times 1°$ horizontal resolution. The Rapid Radiative Transfer Model (RRTM) has been used for long wave (*Mlawer et al.*, 1997) while Dudhia scheme (*Dudhia*, 1989) has been used for short wave radiation. In the model, the sub-grid scale effects of convective and shallow cloud are represented by the cumulus parameterization. The current model set up was tested with Betts-Miller-Janjic (BMJ), Kain-Fritsch (KF) and Grell-Devenyi ensemble (GD) cumulus schemes (Results of KF and GD are not shown). As compared with the observation, the BMJ convective scheme was found to be better and used for the current study. The cumulus parameterization (BMJ scheme) is used in only the outer two domains (d01 & d02). The cloud-resolving 3rd and 4th domain are treated with explicit convection. The microphysical sensitivity of the model was tested with three bulk microphysical parameterization schemes, namely the WRF Double-Moment (WDM6) (*Hong et al.*, 2010), the Thompson scheme (*Thompson et al.*, 2004) and the Morrison double moment with six classes of hydrometeors (*Morrison et al.*, 2005). After a comparison of simulated precipitation and RDSD with the observations (Figure not shown), Morrison double moment scheme was found to be better and hence has been used for all the current simulations. More details regarding the experiment design is documented in Table 3.

Out of the 8 events with stronger in-cloud electrical environment considered for simulation experiment, 5 events were observed over the HACPL,( 17.92 N,73.66 E) which is located in the Western Ghat (WG) of peninsular India at an altitude of 1.3 Km from mean sea level (MSL) with complex topography. All the 5 events with the weaker electric environment were also observed over the HACPL. The pre-monsoon precipitation over the WG is highly convective in nature (*Romatschke and Houze*, 2011) while shallow convective rain dominates the monsoon season (*Konwar et al.*, 2014). The events observed over the HACPL are documented in the Tables 1 & 2 along with some of the available cloud properties and features derived from the Moderate Resolution Imaging Spectroradiometer (MODIS) (Terra platform) collection 6 (*Baum et al.*, 2012) and European Centre for Medium-Range Weather Forecasts (ECMWF) interim reanalysis (ERA-Interim; Dee et al., 2011) at $0.25° \times 0.25°$ resolution datasets.

Table 1: Some Cloud and Electrical properties of the Strongly Electrified (SE) events.

| Dates | Cloud top temperature(K) | Total accumulated rain from JWD (mm) | Total column cloud liquid water (kg m$^{-2}$) | Daily accumulated lightning count in d04 |
|---|---|---|---|---|
| 5 October,2012(a) | 250 | 21.64 | 0.19 | 98 |
| 2 June,2013(b) | - | 30.16 | 0.08 | 186 |
| 10 sept.,2013(c ) | - | 71.5 | 0.04 | 249 |
| 15 May,2015(d) | 252 | 6.98 | 0.002 | 898 |
| 30 May,2015(e) | - | 3.74 | 0.03 | 173 |

**Note**: The lightning counts for the events (a-c) are derived from WWLLN and for (d-e) from MLLN with higher detection efficiency. The total column cloud liquid water was derived from the Era-interim datasets while cloud top temperature was derived from MODIS terra datasets. The labelling for the events is same as Figure 2

Table 2: Some Cloud and Electrical properties of the Weakly Electrified (WE) events.

| Dates | Cloud top temperature(K) | Total accumulated rain JWD (mm) | Total column cloud liquid water (kg m$^{-2}$) | Daily accumulated lightning count in d04 |
|---|---|---|---|---|
| 31Aug,2014(f) | 250 | 116 | 0.75 | 0 |
| 26Oct.,2014(g) | - | 5.4 | 0.06 | 0 |
| 14Nov.,2014(h) | - | 13.41 | 0.01 | 0 |
| 2 Oct.,2015(i) | - | 22.7 | 0.01 | 0 |
| 3 Oct.,2015(j) | 270 | 70 | 0.004 | 0 |

**Note**: The lightning counts for the events (a-e) derived from MLLN. The total column cloud liquid water was derived from the Era-interim datasets while cloud top temperature was derived from MODIS terra datasets. The labelling for the events is same as Figure 2.

The other 3 events with the stronger in-cloud electrical environment were observed over Solapur (17.72$^o$N, 75.85$^o$E) in the rain shadow of the Western Ghat. The observations over Solapur were made in a ground campaign conducted during the Cloud-Aerosol Interaction and Precipitation Enhancement Experiment (CAIPEEX) (*Kulkarni et al., 2012*).

The distinction between stronger/weaker electrical environments is ascertained by the presence/absence of lightning discharges in the innermost domain (Figure 1). The spatial distributions of lightning discharges observed over the model domain d04 for all the events over the HACPL are shown in Figure 2.

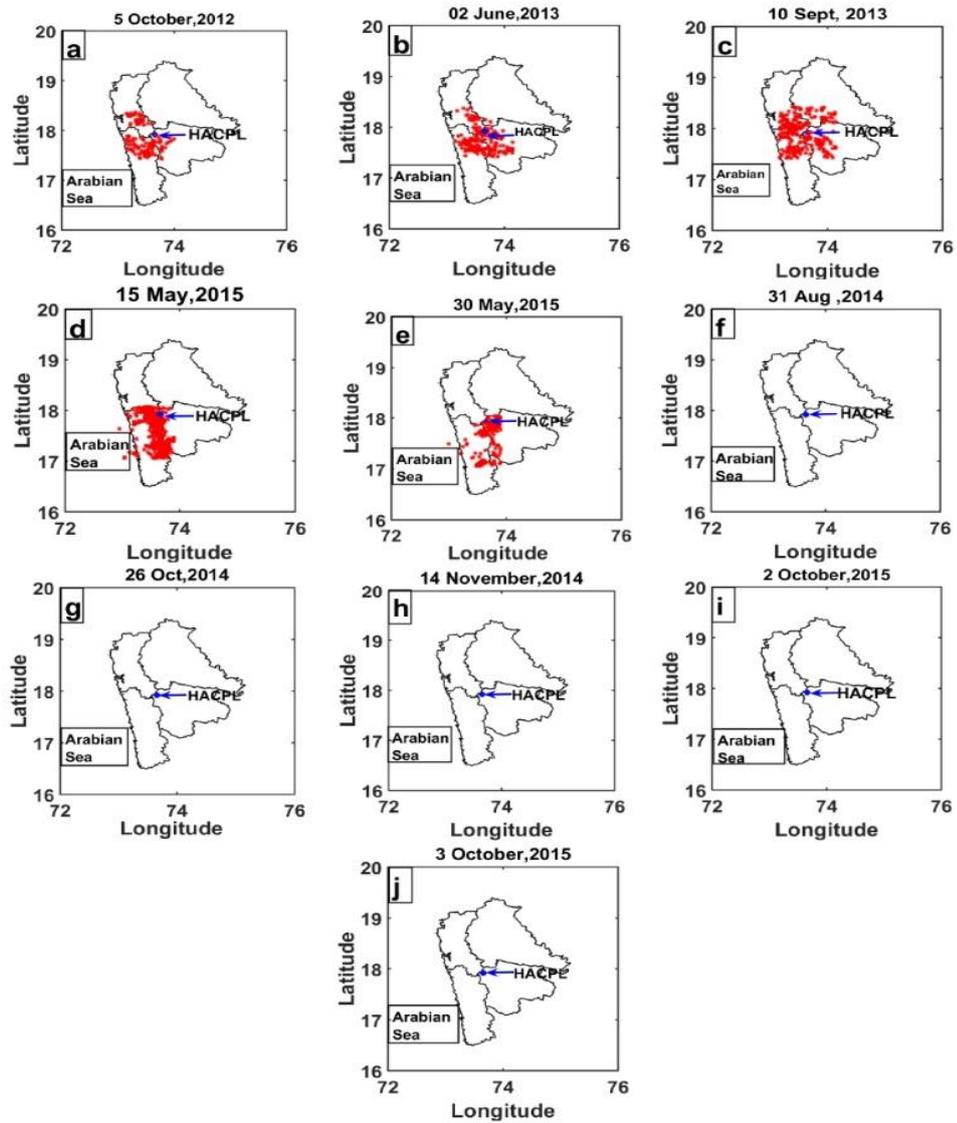

**Figure 2**: The spatial distribution of lightning observed in the model domain d04. Panels (a-e) correspond to strongly electrified (SE) events and (f-j) corresponds to weakly electrified events (WE). The labelling of all the events is same as Table 1 & 2. For events 2(a-c) distribution was derived from the World Wide Lightning Location Network (WWLLN) and for the rest from the Maharashtra Lightning Location Network (MLLN).

While Figures 2(a-e) shows lightning discharges in the area encompassing the model innermost domain, they are conspicuous by absence in the events shown in Figures 2(f-j). For the events (a-c) listed in Table 1, lightning data were extracted from the World Wide Lightning Location Network (WWLLN) with detection efficiency of 25% -30% while for the rest; they were extracted from Maharashtra Lightning Location Network (MLLN) (*Pawar et*

*al*., 2017) with detection efficiency about 90%. The MLLN operates in the frequency of 1 KHz (VLF) and 10 MHz (HF). As lightning-producing clouds exhibit stronger electrical environment in terms of magnitude of electric field and charge distribution, we termed these set of events as strongly electrified (SE) events while non-lightning-producing rain events were termed as weakly electrified (WE) events.

The experiments were carried out as discussed below

A set of control (CTL) experiments were carried out for both the SE and WE of events with WRF-ARW model with the standard physics packages using the same model set up and the simulated precipitation field and RDSD were validated against available observed variables.

In the WRF-ARW, the precipitation is calculated using a Marshall-Palmer formulation of RDSD with a specified slope parameter, $\lambda$. We find that $\lambda$, used in the default physics scheme is inadequate to represent precipitation in the SE events. Hence in a second set of experiment, the default minimum value of the RDSD slope parameter, $\lambda$ in the physics module has been replaced with a new $\lambda$ as obtained from observation, averaged over all the five SE events observed over the HACPL. The influence of $\lambda$ on the simulated precipitation has been discussed in the supporting text in details. A new set of simulations was carried out for the same SE events using the same model setup with the modified physics.

For the rain events recorded in the afternoon or late afternoon hours, the model was initialized with the NCEP FNL 00:00:00 UTC initial conditions (IC) while for the late night or early morning events, initialization was done using the 12:00:00 UTC ICs. The details of the model design have been tabulated in Table 3.

**Table 3: The WRF Model Experiment Design.**

|  | Name of Experiment | | | |
|---|---|---|---|---|
| **Physical Processes** | Control (CTL) run | Modification of limit of $\lambda$ in Morrison scheme | Modification of aerosol number concentration | Modification of aerosol number concentration + Modification of $\lambda$ |
| **Convective process** | Betts-Miller-Janjic (BMJ) | Betts-Miller-Janjic (BMJ) | BMJ | BMJ |
| **Microphysics process** | Default Morrison Scheme (Morr) following *Morrison et al.*,2005. | Modified Morrison (Morr(M)). The default minimum value of $\lambda$ in the physics module has been replaced with a new $\lambda$, averaged over all the five SE events observed over the HACPL | Default Morrison Scheme (Morr) + change in aerosol number concentration in Mode 1(0.05 µm) | Modified Morrison (Morr(M)) + change in aerosol number concentration in Mode 1(0.05 µm) |
| **Model Initialization** | For the events (b-d) documented in the Table 1, (a,c,d,e) in Table 2 and the events (a-b) in Figure 6, the model was initialized with 00:00:00 UTC NCEP ICs while for the events ( a & e) in Table 1, (b) in Table 2 and event (c) in Figure 6 , model was initialized with 12:00:00 UTC IC. | | | |

**Note**: Other physical processes (short and long wave radiation scheme) are kept same for both sets of sensitivity experiment.

For comparison with observations, data from surface-based JW disdrometer (JWD) located at the HACPL and Solapur were used which record the RDSD and rain intensity (Joss and Waldvogel, 1967). The hourly accumulated rain was extracted from the JWD record and considered for validation of simulated hourly precipitation. The recorded distribution was also used to calculate the RDSD parameters from the gamma distribution fitted to RDSD. Data recorded by an optical disdrometer installed at Pune (18.52$^o$N, 73.85$^o$E) was also used to study the geographical variability of the RDSD. Aerosol distribution was observed over the HACPL with a Scanning Mobility Particle Sizer (SMPS) while Cloud Condensation Nuclei (CCN) was measured with a collocated Cloud Condensation Nuclei Counter (CCNC) (*Singla et al.*,2019).

## 3. Comparison of Simulations with Observation

### 3.1 Accumulated Precipitation and RDSD

#### 3.1.1 Strongly Electrified

The reported underestimation of simulated rain intensity for the rain events that are associated with lightning discharges suggested model's inability to reproduce heavy precipitation amount towards higher rain bins. The underestimations of rainfall are linked with the improper representation of RDSD. In the present study, the 5 SE events observed over the HACPL are simulated and verified for precipitation comparing with the JWD measured hourly rain rate. As we have not addressed the spatial displacement of the simulated center of mesoscale convection relative to the observation, the simulated precipitation is verified in all the grid points within a 25km × 25 km box, centered at the HACPL. The grid point that shows the closest value of precipitation rate to the observed one is considered as model simulated precipitation and compared with the observation. Figure 3 (a-e) shows the simulated rain rate for the events reported in Figure 2(a-e) along with the observed rain rate.

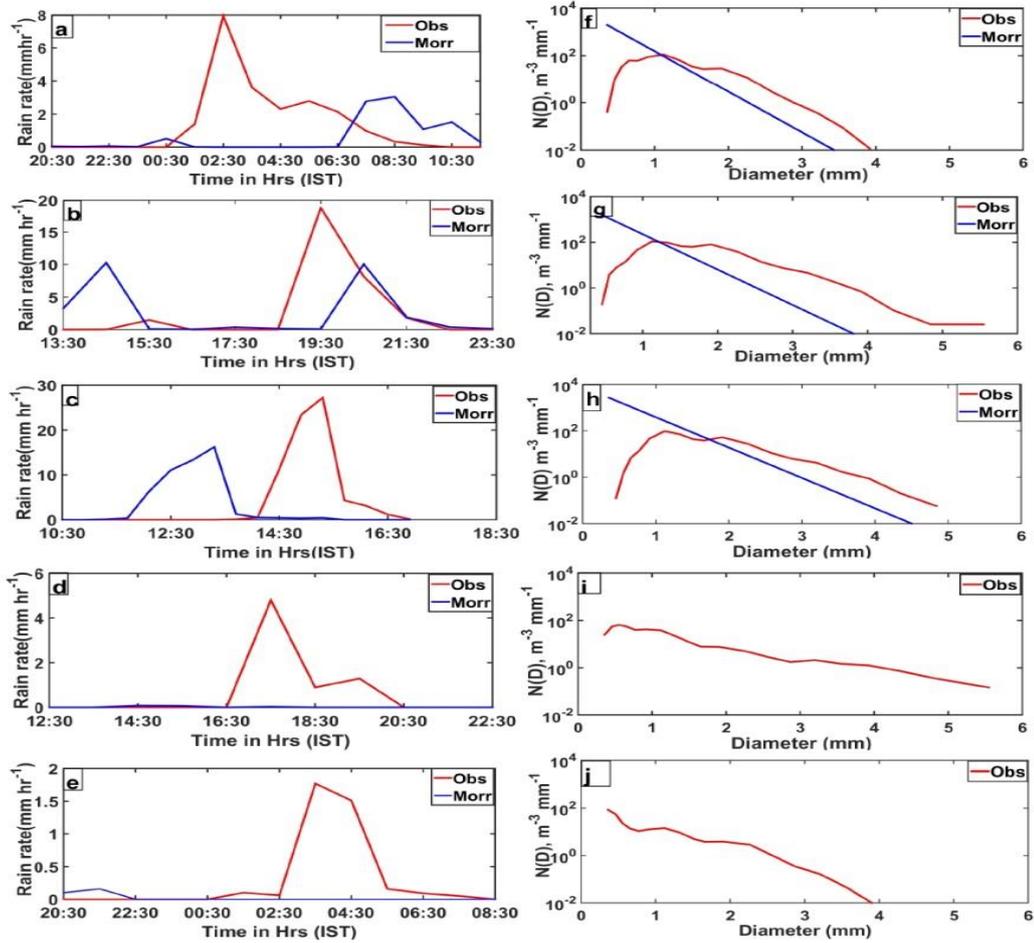

**Figure 3:** Comparison of simulated rain rate (a-e) and simulated RDSD (f-j) with the observation for the strongly electrified (SE) events observed over the HACPL. N(D) is the number density of drops. The legends in the figure 'Obs' indicated observation while 'Morr' indicated Morrison double moment scheme.

Apart from the shift in timing of the peak rainfall, significant underestimation of the observed precipitation can be seen in the simulations for the events 3(a-c) while for events 3(d-e), model failed to simulate any rain during the event duration. The underestimation of rain intensity is found to be consistent with the earlier reported dry bias in the simulation of heavy precipitation associated with lightning activity (*Giinaros et al.,*2015; *Dafis et al.,* 2018). In some cases, the model predicted the rainfall 3-4 hours advance while in others the rainfall was delayed by 1-2 hours. This phase difference in the diurnal cycle of the simulated peak and observed peak in precipitation is well recognized (*Jeong et al.,* 2011; *Diro et al.,* 2012 ; *Walther et al.,* 2013; *Gao et al.,* 2018).

The higher sensitivity of model accumulated precipitation to the prescribed RDSD was reported in a number of earlier studies (*Gilmore et al.*, 2004; *Curic et al.*, 2010; *Morrision*, 2012; *Kovacevic and Curic*, 2015). The prognostic variables like mixing ratios and number concentrations of different species of hydrometeors are expressed as a function of RDSD parameters. In Figure 3(f-j), the model simulated RDSDs were compared with the observed RDSD. The observed RDSD was averaged over the entire duration of rainfall for each event. The simulated RDSD was calculated using the model predicted rain mixing ratio averaged over the rain period. The double moment microphysics scheme predicts the mass mixing ratios and number concentration of hydrometers assuming gamma particle size distribution

$$N(D) = N_0 D^\mu e^{-\lambda D} \qquad (1)$$

Where $N_0$, $\lambda$, $\mu$ are the intercept, slope and shape parameters of the size distribution, respectively. D is the diameter of the particles.

With µ=0 for rain (*Morrison et al.*, 2008), the size distribution of rain will take the form of exponential functions (Marshall-Palmer distribution)

$$N(D) = N_0 e^{-\lambda D} \qquad (2)$$

$\lambda$ & $N_0$ can be derived from the model predicted rain number concentration N and rain mixing ratio q

$$\lambda = \left(\frac{\pi \rho_r N}{q\rho}\right)^{1/4} \qquad (3)$$

$$N_0 = N \lambda \qquad (4)$$

Where $\rho_r$ is the density of raindrops (1000 kg m$^{-3}$) and $\rho$ is the air density.

Consistent with the underestimation of observed rainfall intensity by the model in the events shown in Figure 3(a-c), the simulated RDSD in Figure 3(f-h) shows substantial underestimation in the number concentration of larger raindrops compared to the observation. As the model was unable to reproduce rainfall at the surface for the SE events shown in Figure 3(d-e), the RDSD corresponding to these events only depicts the observed distributions (Figure 3(i-j)). The overestimation of the smaller-size raindrops may be caused by the inherent deficiency of assumed Marshall-Palmer distribution (*Gao et al.*, 2018). It was observed that the underestimation in drops number concentration increases as drop size increases.

### 3.1.2 Weakly Electrified

It is interesting to note that there is no underestimation of observed rainfall by the model in the WE cases (Figure 4(a-e)).

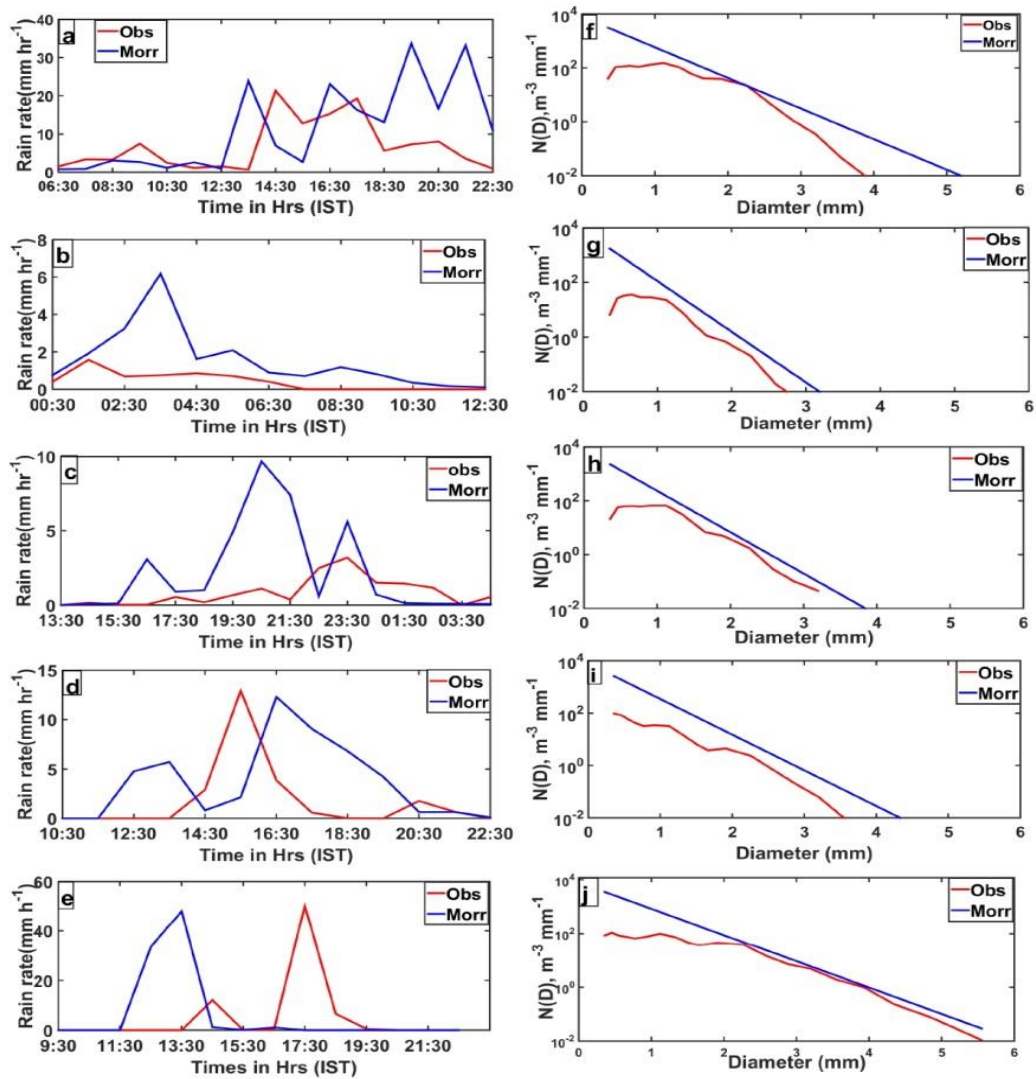

**Figure 4:** As in Figure 3 but for weakly electrified (WE) events.

Apart from the generic problem of timing of peak rainfall simulation, the model in fact slightly overestimated the precipitation intensity compared with the observation in three out of five events as shown in Figure 4(a-c). This wet bias in the WE events was found to be in contrast with the reported dry bias in SE events. The temporal spread in the simulated rain was found to be consistent with the observation. For some of the events, the phase shift in the precipitation peak was found to be 1-4 hours.

The right panels of Figure 4(f-j) depict the comparison of the simulated RDSD with the observed ones. Both the sets of RDSD are averaged over the entire rain duration recorded by the model and JWD. The observed RDSD for the WE events primarily found to be of exponential in nature and comparable with the simulated ones in almost all the events, as shown in Figure 4(f-j). It is also observed that in both types of events, the model overestimated the number concentration of smaller size drops. In the case of WE events for higher rain intensity, the tail of the distribution was extended towards the larger drop size in the JWD measured RDSD, while in contrast, broadening of the RDSD towards the larger size range was observed irrespective of the rain intensity for SE events.

Thus, for the WE events, the simulated precipitation and the RDSDs were found to be comparable with the observations while for the SE cases intensity of rainfall is underestimated consistent with significant underestimation of larger drops indicating a potential limitation in the RDSD parameterization in the Morrison microphysics used in the WRF-ARW model. An appropriately modified Morrison scheme for SE cases and re-simulation of the SE cases with the modified scheme is presented next.

### 4. Modification of Model RDSD for Strongly Electrified events

It is clear that the inability of the model to simulate the intensity of precipitation in the SE cases is related to its bias in simulating the larger drops in the RDSD. The fact that the slope of the RDSD in the WE cases match well with that of observations, indicates that the model specification of 'slope' for SE cases is inadequate. Modeling studies (*Gilmore et al.*, 2004; *Curic et al.*, 2010; *Morrison*, 2012; *Kovacevic and Curic*, 2015) indicating that the simulated precipitation is sensitive to the prescribed RDSD parameters viz. µ, λ and $N_o$ as they are explicitly depends on the prevailing microphysical processes (*Konwar et al.*, 2014). By virtue of microphysical modification of λ through enhanced collision-coalescence growth of raindrops in presence of stronger in-cloud electric forces (*Khain et al.*, 2004; *Harrison et al.*,2015*; Mudiar et al*., 2018), the characteristic value of λ for the SE events is expected to be distinct from the WE one. Here we demonstrate that the simulated precipitation exhibits significant improvement if modification of the RDSD by electrical effect is adequately included for the SE events. The modification of physics is primarily through modification of the slope parameter λ (mm$^{-1}$) that are estimated using moments method reported in *Konwar et al.*,(2014) and documented in Table 4.

**Table 4**: The RDSD parameters for the five events reported in figure 2(a-e) obtained from surface-based JW disdrometer for the SE events over the HACPL.

| Date | μ | $\lambda$ (mm$^{-1}$) | $N_o$ (m$^{-3}$ mm$^{-(\mu+1)}$) | Mean λ | S.D. of λ |
|---|---|---|---|---|---|
| 5 October,2012(a) | 5.1 | 4.88 | 6.2×10$^3$ | | |
| 2 June,2013(b) | 3.44 | 3.26 | 1.8×10$^3$ | | |
| 10 sept.,2013(c ) | 5.2 | 3.95 | 2.9×10$^3$ | 3.31 | 1.28 |
| 15 May,2015(d) | 0.45 | 1.39 | 70 | | |
| 30 May,2015(e) | 2.1 | 3.10 | 390 | | |

**Note:** μ is the shape parameter, $\lambda$ slope parameter (mm$^{-1}$) and $N_o$ intercept parameter (m$^{-3}$ mm$^{-(\mu+1)}$). RDSD parameters are estimated using the M234 method following *Konwar et al.* (2014). Labelling of the events is same as Table1.

As indicated in the physics module of WRF (Morrison), earlier attempt has been made to increase the minimum value of λ for rain in the WRF version 3.2, although as would be seen from the current study, use of a universal λ may be responsible for the observed discrepancy between simulated and observed rainfall in the case of the SE and WE events. In the sensitivity experiment, the default minimum value of the λ in the physics module has been replaced with a new λ averaged over all the five SE events as obtained from observation over the Indian subcontinent. The modified simulated precipitation is shown in green colors in Figures 5(a-e) indicated as 'Morr(M)' along with the default Morrison indicated as 'Morr' together with the observed ('Obs').

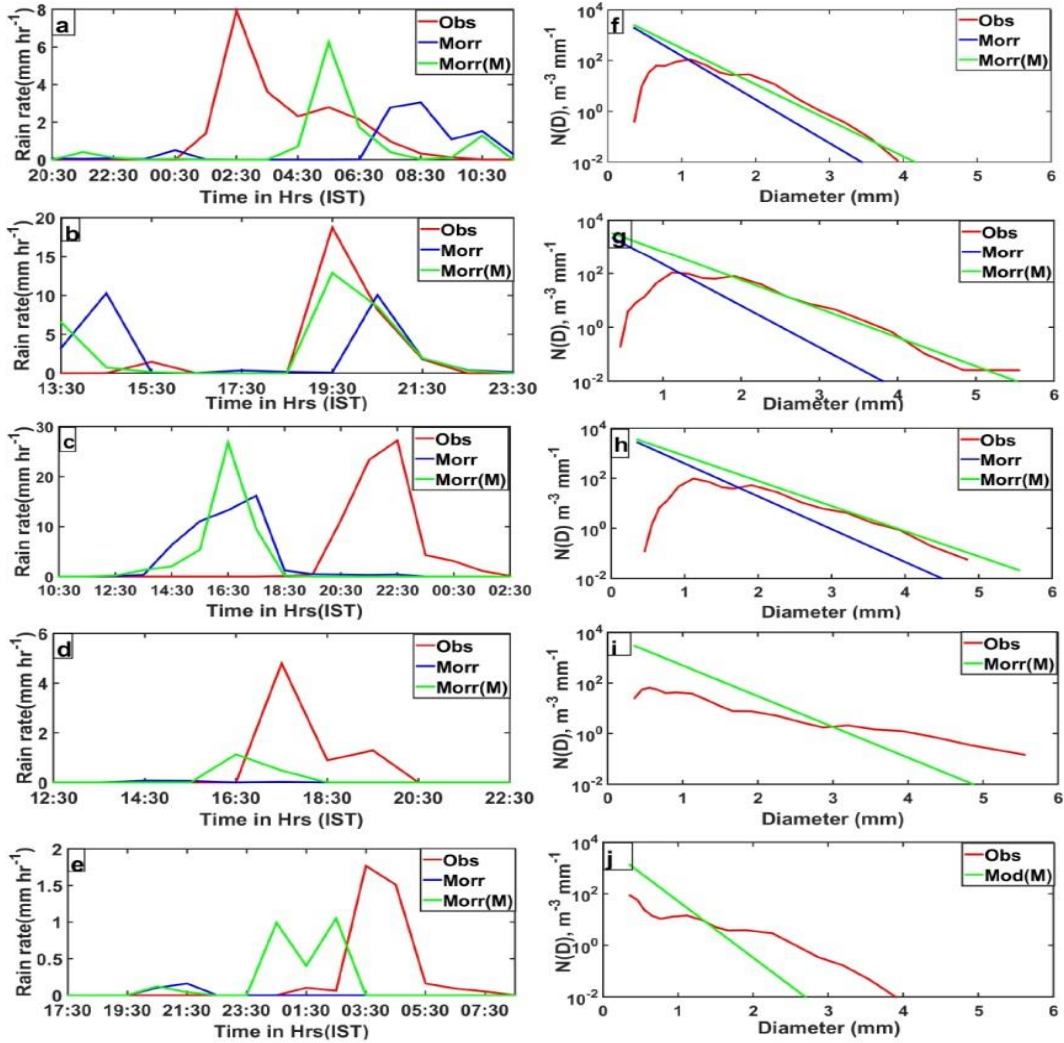

**Figure 5**: As in Figure 3 but with modified Morrison scheme. The legends 'Obs' indicated observation, 'Morr' indicated Morrison scheme and 'Morr(M)' indicated modified Morrison scheme.

Substantial improvement was observed in rain intensity with the incorporation of electrically modified λ in all the events. For the events shown in Figure 5(d-e), for which the default Morrison scheme was unable to reproduce any rain for the simulated period, the model with the Morr(M) reproduces substantial amount of rain albeit with some underestimation still remaining. The right panels of Figure 5 depict the simulated RDSD with the modified scheme along with the default and the observed ones. Substantial improvement in number concentrations of larger raindrops can be observed with the Morr(M) ( Figure 5(f-h)). While for the events shown in Figure 5(i-j), the simulated RDSD show some improvement consistent with the larger amount of simulated rainfall, with underestimation of larger raindrops still persisting. The overestimation of the number concentration of the smaller size drops still persists. The overall improvement in the accumulated rain and RDSD indicate considerable sensitivity of simulated precipitation to λ and establishes the benefit of electrically modified slope parameter, λ.

In order to ascertain the representativeness of λ derived over the HACPL, we have investigated the spatio-temporal variability of λ for SE events. For this purpose, we have evaluated λ considering some additional SE rain events (other than the events documented in Table 1) associated with lightning over the HACPL as well other two locations in the state of Maharashtra, e.g., in Pune & Solapur. While the HACPL is located in the windward slope of the WG, Pune and Solapur are located in the leeward side of WG with MSL height of 560m and 458m respectively. The values are documented in Table S1, S2 and S3 and found to be in a similar range as in Table 4. Figure 6 depicts the results of simulation of 3 SE events observed over Solapur using the same microphysical and cumulus schemes.

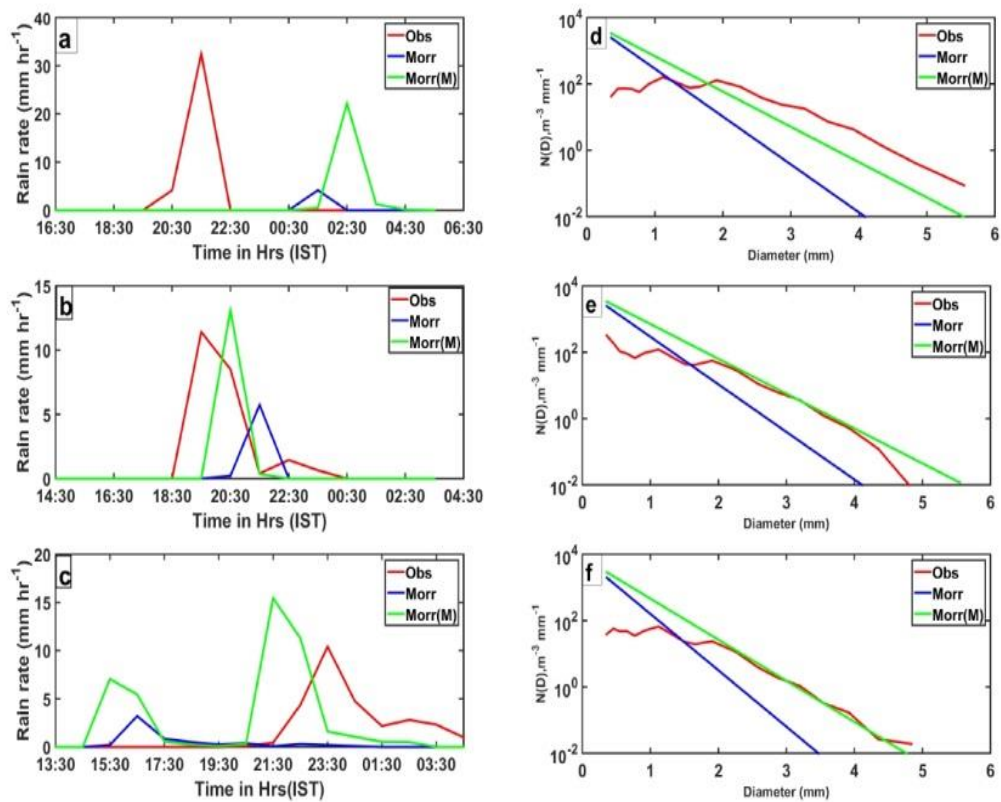

**Figure 6:** Comparison of simulated precipitation (a-c) and corresponding RDSD (d-f) with the observation for the SE events observed over Solapur documented in table S3. The legends 'Obs' indicated observation, 'Morr' indicated Morrison scheme and 'Morr(M)' indicated modified Morrison scheme.

The modified simulated precipitation (Morr(M)) corresponds to the same value of λ as over the HACPL. The substantial improvement in the precipitation field as well as in the RDSD with the modified physics over the HACPL as well as over Solapur, a region of significantly lower climatological mean rainfall added confidence to our conclusion that the effect of electrically enhanced coalescence growth of raindrops in precipitation formation inside SE cloud is valid irrespective of geographical locations. Table S4 depicts a few representative values of λ for the WE events over the HACPL, distinguishable from the values in the SE ones with higher magnitude & variability.

### 5. Aerosol and CCN influences on the simulated Rain Intensity:

One factor that could add a certain amount of uncertainty to our primary conclusion is the precipitation modification by aerosol concentrations. Numerous in-depth studies reported significant modification of accumulated precipitation by ambient aerosol concentration, although the relationship between the two observables is quite non-linear (*Khain et al*. 1999; *Rosenfeld*, 2000; *Rosenfeld et al.* 2002; *Rosenfeld and Woodley*, 2003; *Andreae et al*., 2004; *Khain et al*., 2005). To get more confidence in our primary hypothesis of electrical enhancement of precipitation intensity, we further investigated the response of precipitation to the number concentration of aerosol, which can act as Cloud Condensation Nuclei (CCN) over the HACPL. Aerosol distribution was measured over the HACPL with Scanning Mobility Particle Sizer (SMPS) while Cloud Condensation Nuclei (CCN) was measured with a collocated Cloud Condensation Nuclei Counter (CCNC) (*Singla et al*., 2019). The measured CCN number concentration over the HACPL indicated no discernible difference between the SE and WE events (Figure 7a).

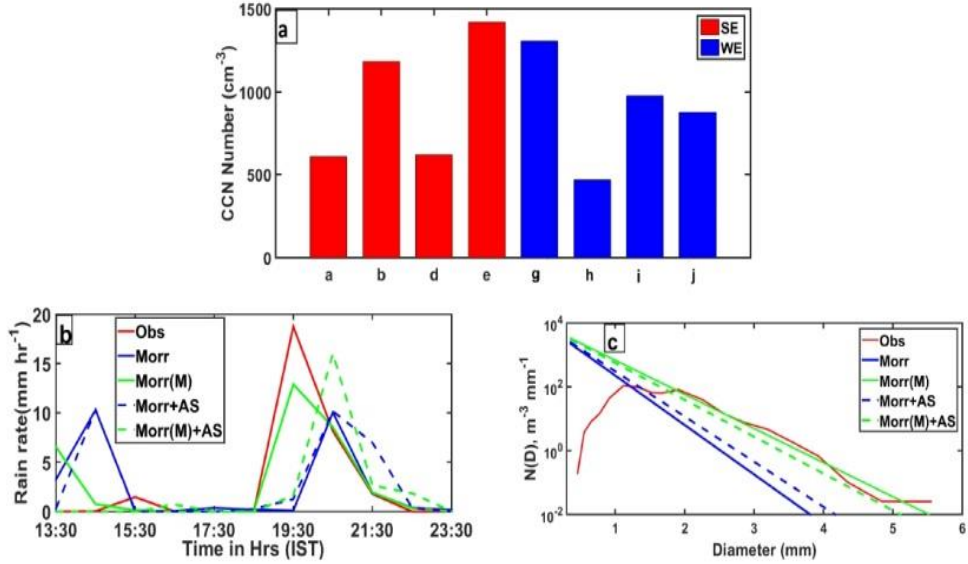

**Figure 7**: (a) Bar representation of total CCN number concentration for SE and WE events at 0.3% supersaturation. Labelling is same as Table 1& 2. (b) Comparison of simulated rain intensity with aerosol modification. (c) Comparison of RDSD. The simulation with aerosol modification alone is indicated as 'Morr+AS' while simulation with both aerosol and λ modification is indicated as 'Morr(M)+AS' while 'Morr' indicated Morrison scheme and 'Morr(M)' indicated modified Morrison scheme.

The measured aerosol concentration shows little higher value for SE (157 cm$^{-3}$) events compared to that of WE (137 cm$^{-3}$), both exhibiting peak around 0.05 μm (*Aitken mode*). When the model physics is perturbed by adding the difference between the mean concentration of aerosol and the one observed for SE events (around 6% of mean), it is observed that increase of aerosol concentration alone doesn't substantially change the simulated intensity of precipitation, although adds a little to the total accumulation (Figure 7b). However, when the aerosol perturbation is added with Morr(M), simulated intensity shows a discernible improvement while the peak rainfall is delayed by an hour. This delaying is expected as a higher concentration of aerosol would reduce the drop sizes inhibiting collision-coalescence growth of drops in the warm phase, thereby suppressing the warm rain by the first aerosol indirect effect (*Twomey et al*., 1984, *Hazra et al*., 2013). The RDSDs shown in Figure 7(c) also do not indicate a significant change in the number concentration of larger drops by aerosol modification, although the modification through electrically modified λ is quite significant. Results of this experiment (see supporting text for details) adds to our confidence on the primary conclusion of electrical modification of simulated precipitation intensity.

## 6. Discussions and Conclusion

In quest of better estimation of precipitation for the benefit of meteorological as well as hydrological applications and encouraged by compelling evidences from laboratory as well as field experiments on substantial influences of electrical forces on cloud/rain microphysical processes, here we demonstrate that modelling the RDSD correctly in an NWP model is critical in simulating and predicting the rainfall with fidelity in MCS. A microphysical modification in the model emerged from a set of simulations of 8 SE cases and 5 WE cases with Morrison microphysics and critical evaluation of their biases. The results suggested that the underestimations of heavy rainfall associated with SE events may be caused by the model's inability to properly reproduce the larger raindrops which get substantially improved with the inclusion of electrically modulated RDSD slope parameter $\lambda$. The vertical profiles of simulated hydrometeors presented in Figure 8 suggested that the two sets of event (SE and WE) observed over the HACPL which are selected for the testing of our primary hypothesis are similar to each other, apart from their electrical distinguishability.

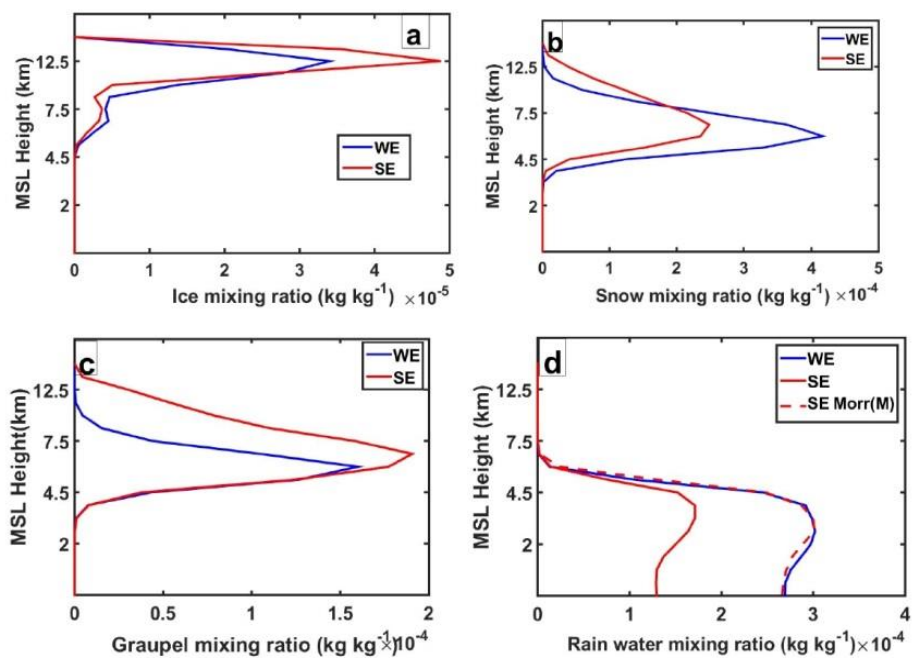

**Figure** 8: Area and time averaged vertical distribution of simulated (a) Ice mixing ratio (kg kg$^{-1}$), (b) Snow mixing ratio (kg kg$^{-1}$), (c) Graupel mixing ratio (kg kg$^{-1}$) (d) Rain mixing

ratio (kg kg$^{-1}$) for the events observed over the HACPL. The blue and red curves corresponds to the WE and SE events respectively. Each profile has been averaged over 5 events each.

A little higher cloud ice and graupel mixing ratio (Figure 8a & 8c) are expected for the lightning-producing clouds required by non-inductive charging mechanism of cloud electrification. It is also seen that in case WE events, graupel resides at lower altitude in the mixed phase region of cloud compared to the SE events (supported by observation of *Mattos et al*., 2016). This suggests the presence of larger size graupel particles above the melting layer in WE events relative to the SE ones, where they are more numerous. The similar vertical distribution of rain mixing ratio (Figure 8d) (which shows a substantial increase with the inclusion of modified slope parameter) indicated a similar cloud environment for both sets of events. The underestimation in rain intensity in SE events despite having similar (or higher) graupel and ice mass as WE events suggested that the underestimation in the observed rain intensity may primarily cause by some missing microphysical processes influenced by electric forces, which broaden the RDSD and hence enhance the growth rate of raindrops. The melting of larger graupels although can contribute to the total rain, the result suggests it is not contributing much to the biases in intensity simulation of SE events. The improvement in rain intensity with inclusion of characteristics slope parameter which observed to be distinguishable between SE and WE events (Table 4, S1, S2, S3 and S4) suggested a substantial influences of electrical forces (possibly in the warm phase of the cloud by virtue of enhanced collision-coalescence growth of raindrops) on the rain formation processes inside the SE cloud.

The simulations presented in the study are short-range predictions and as such sensitive to initial conditions (IC). The coarse resolution analysis (NCEP-FNL) interpolated to the finer model domain may introduce some uncertainty in the IC over the finer resolution model domain. To test the robustness of our main result that Morr(M) improves the simulation of rainfall intensity and RDSD in the SE cases, we carried out an ensemble of simulations adding a 'small' perturbation in the temperature field of the NCEP ICs (see Supporting text for details) for the event (b) documented in the Table 1. It was observed (Figure 9) that the rain intensity as well the RDSD does not show significant sensitivity to the perturbed ICs while the sensitivity to the electrically modified $\lambda$ is highly significant suggesting the robustness of our primary conclusion.

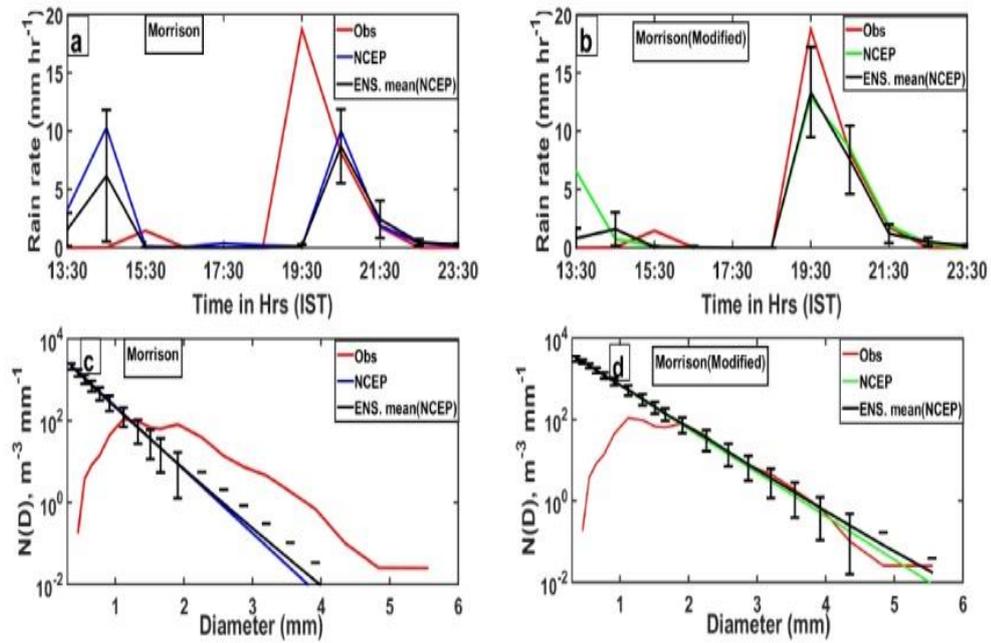

**Figure 9**: Inter-comparison among observation and simulation with NCEP Initial Condition (IC) as well with ensemble mean of 10 member ensemble generated by perturbing the temperature field of NCEP IC, labelled as Obs, NCEP, ENS. Mean (NCEP) respectively. (a-b) precipitation intensity, (c-d) RDSD. The vertical bars indicate the respective standard deviation.

We are aware that this is primarily a sensitivity study subject to knowledge of the observed RDSD. However, substantial improvement in the simulated precipitation with the electrically modulated RDSD parameters provides a promising pathway for parameterizing the electrical forces in weather/climate models. The optimism is based on our recent findings (Mudiar et al., 2018) that quantify the modification of RDSD by electrical forces in stratiform as well as convective rain events. With the parameterization of the electrical effect in the physics module of NWP model, the reported dry bias associated with heavy precipitation events in the weather/climate models is likely to be minimized and increase the skill of the models in predicting intensity of quantitative precipitation.

**Acknowledgments**

IITM is funded by the Ministry of Earth Sciences, Government of India. We thank Director, IITM for his continuous support and encouragement. We sincerely acknowledge the efforts of scientific and technical staffs of IITM, Pune, working in the HACPL to collect the JWD data. First author is grateful to Madhuparna Halder for the support in initial model setup. The data used to prepare the manuscript is available in the link https://iitmcloud.tropmet.res.in/index.php/s/g87WJjyayHLw42n. We are thankful to the University of Washington in Seattle, USA for making available the WWLLN data (http://wwlln.net/). BNG thanks Science and Engineering Research Board, Government of India for the SERB Distinguished Fellowship. Authors are grateful to CAIPEEX team members for their efforts in collecting data.